\documentclass[useAMS,usenatbib]{mn2e}
\usepackage{txfonts,graphicx,natbib}
\bibpunct{(}{)}{;}{a}{}{,}

\def \aj {AJ}
\def \mnras {MNRAS}
\def \pasp {PASP}
\def \apj {ApJ}
\def \apjs {ApJS}
\def \apjl {ApJL}
\def \aap {A\&A}
\def \nat {Nature}

\def \aaps {A\&A Suppl.}
\def\lesssim{\mathrel{\hbox{\rlap{\hbox{\lower4pt\hbox{$\sim$}}}\hbox{$<$}}}}
\def\gtrsim{\mathrel{\hbox{\rlap{\hbox{\lower4pt\hbox{$\sim$}}}\hbox{$>$}}}}

\DeclareMathAlphabet{\mathsc}{OT1}{cmr}{m}{sc}
\def\testbx{bx}%
\DeclareRobustCommand{\ion}[2]{%
\relax\ifmmode
\ifx\testbx\f@series
{\mathbf{#1\,\mathsc{#2}}}\else
{\mathrm{#1\,\mathsc{#2}}}\fi
\else\textup{#1\,{\mdseries\textsc{#2}}}%
\fi}

\newcommand{\Hii} {\ion{H}{ii}}

\begin{document}
\title[The progenitor of SN 2008ax]{The type IIb SN 2008ax: the nature of the progenitor}
\author[Crockett et al. ]
{R. M. Crockett,$^{1}$\thanks{E-mail: rcrockett02@qub.ac.uk} J. J. Eldridge,$^{2}$ S. J. Smartt,$^{1}$ A. Pastorello,$^{1}$ A. Gal-Yam,$^{3}$
\newauthor D. B. Fox,$^{4}$ D. C. Leonard,$^{5}$ M. M. Kasliwal,$^{6,7}$ S. Mattila,$^{8}$ J. R. Maund,$^{9}$
\newauthor A. W. Stephens,$^{10}$ I. J. Danziger$^{11,12}$\\
$^{1}$Astrophysics Research Centre, School of Maths and Physics, Queen's University Belfast, Belfast BT7 1NN, UK\\
$^{2}$Institute of Astronomy, The Observatories, University of Cambridge, Madingly Road, Cambridge CB3 0HA, UK\\
$^{3}$Benoziyo Center for Astrophysics, Weizmann Institute of Science, 76100 Rehovot, Israel\\
$^{4}$Department of Astronomy and Astrophysics, Pennsylvania State University, University Park, PA, US\\
$^{5}$Department of Astronomy, San Diego State University, San Diego, CA 92182, US\\
$^{6}$Division of Physics, Mathematics, and Astronomy, 105-24, California Institute of Technology, Pasadena, CA 91125, US\\
$^{7}$Hale Fellow, William and Betty Moore Foundation\\
$^{8}$Tuorla Observatory, University of Turku, V\"ais\"al\"antie 20, FI-21500 Piikki\"o, Finland\\
$^{9}$Department of Astronomy and McDonald Observatory, University of Texas, 1 University Station, C1400, Austin, TX 78712, US\\
$^{10}$Gemini Observatory, 670 North A'ohoku Place, Hilo, HI 96720, US\\
$^{11}$INAF, Osservatorio Astronomico di Trieste, via G.B. Tiepolo 11, 34131 Trieste, Italy\\
$^{12}$Department of Astronomy, University of Trieste, via G.B. Tiepolo 11, 34131 Trieste, Italy\\
}
\maketitle
\label{firstpage}
\begin{abstract}
A source coincident with the position of the type IIb supernova (SN) 2008ax is identified in pre-explosion HST WFPC2 observations in three optical filters. 
We identify and constrain two possible progenitor systems: (1) a single massive star that lost most of its hydrogen envelope through radiatively driven mass loss processes, prior to exploding as a helium-rich Wolf-Rayet star with a residual hydrogen envelope, and (2) an interacting binary in a low mass cluster producing a stripped progenitor. Late time, high resolution observations along with detailed modelling of the SN will be required to reveal the true nature of this progenitor star.

\end{abstract}

\section{Introduction}
\label{intro}

Although the progenitors of several type II supernovae (SNe) have been identified in pre-explosion observations \citep[e.g.,][]{smartt04,li06}, the stripped stars thought to explode as type Ib/c SNe have so far eluded discovery \citep[e.g.,][]{maundsmartt05,maund05,galyam05,rmc07,rmc08}. The stars exploding as type IIb SNe (those events which transition from type II to type Ib) are believed to be intermediate of the hydrogen-rich (H-rich) supergiant type II and the H-free type Ib precursors.  The progenitor of the nearby type IIb SN 1993J was shown to be such a star; a K0 supergiant stripped of most of its hydrogen envelope by a massive binary companion \citep{ald94,maund04}. 

In this letter we report the identification and characterisation of the progenitor of the
type IIb SN~2008ax, in the galaxy NGC~4490.  SN 2008ax was discovered by \citet{most08} on 2008 Mar 3.45 at $\mathrm{\alpha_{2000}=12^{h}30^{m}40.^{s}80}$, $\mathrm{\delta_{2000}=41\degr38\arcmin14.\arcsec5}$, just 6 hrs after its non-detection in an unfiltered observation made by R.~Arbour with a limiting magnitude of 18.5 \citep{nak08}. SN~2008ax was spectroscopically classified as a type IIb SN by \citet{chorn08}. We have obtained extensive and high quality spectro-photometric monitoring of this SN, the analysis of which is presented in a companion paper \citep{past08}.

\section{Observations and Data Reduction}

The pre- and post-explosion observations analysed in this study are summarised in
Table \ref{obstab}.  Archival Hubble Space Telescope (HST) images of the 
site of SN 2008ax prior to explosion were recovered from the STScI archive\footnote{http://archive.stsci.edu/hst/} and calibrated via the on-the-fly-recalibration (OTFR) pipeline.  These observations were taken using the Wide Field Planetary Camera 2 (WFPC2) through four optical filters on three different epochs.  The site of SN 2008ax was located on the WF3 chip of the F450W and F814W frames, the WF2 chip of the F606W image and the PC chip of the F300W image. The WF chips have pixel scales of $\mathrm{0.1\arcsec~pix^{-1}}$ while the PC chip pixel scale is $\mathrm{0.046\arcsec~pix^{-1}}$.  Each of the WFPC2 observations were taken as two exposures which were combined in order to remove cosmic rays. Pixel offsets noticed between the F606W exposures were taken into account during the combination process. Finally the combined images where corrected for geometric distortion.  These calibrated images were used to perform the image alignment described in \S 3. PSF photometry was carried out using the {\sc hstphot} package (version 1.1.7b) \citep{dolhstphot}. Option 10 was chosen, which turned on local sky determination and turned off aperture corrections since there were no good aperture stars. (In this case {\sc hstphot} applies default filter-dependent aperture corrections). The F606W exposures were input as individual frames along with appropriate offsets, since the {\sc hstphot} pre-processing task {\it coadd} is unable to apply shifts before combining images. Our photometry in this filter is $\sim$0.3 mag brighter than in \citet{li08b} who may have combined the exposures without applying offsets, thereby clipping the stellar profiles.  

Ground-based adaptive optics observations of SN~2008ax where taken on 2008 April 3 using Altair/NIRI on the 8.1-m Gemini (North) Telescope. These observations were carried out as part of programs GN-2008A-Q-28 (PI: Crockett) and GN-2008A-Q-26 (PI: Gal-Yam).  Short exposures (to keep the SN counts linear) totalling 1200 sec on-source were taken in the $K$-band using the SN as the natural guide star (NGS). The images were reduced, sky-subtracted and combined using the NIRI reduction tools within the {\sc Iraf} {\it gemini} package.  The final reduced image is of very high quality showing near-diffraction limited resolution of $0.09\arcsec$ and Strehl ratios of up to $\sim$10 percent (Figure 1). The pixel scale of the NIRI f/32 camera was $\mathrm{0.022\arcsec}$ providing excellent sampling of the PSF.

\label{obs}
\begin{table}
\caption{\label{obstab} Pre-and post-explosion observations of the site of SN 2008ax.}
\begin{tabular}{lrrc}
\hline\hline
Date  & Telescope/Instrument & Filter  & Exp. Time \\
      &            &         & (s)       \\
\hline
\multicolumn{4}{c}{{\bf Pre-explosion images}} \\
2001 Jul 2 & HST/WFPC2 & F300W & 600\\
2001 Nov 13 & HST/WFPC2 & F450W & 460 \\
1994 Dec 3 & HST/WFPC2 & F606W & 160 \\
2001 Nov 13 & HST/WPFC2 & F814W & 460 \\
~\\\hline
\multicolumn{4}{c}{{\bf Post-explosion images}} \\
2008 April 3 & Gemini-N/Altair+NIRI & K & 1200 \\
\hline\hline
\end{tabular}
\end{table}

\section{Analysis and Results}

A transformation between the pre- and post-explosion coordinate frames was derived in order to identify the precise position of the SN on the archival HST images. The positions of 35 point sources common to both the HST F814W and the Gemini K-band frames were input to the {\sc Iraf} task {\it geomap}, which calculated a general geometric transformation with a rms error of $\pm$22 mas.  The position of the SN in the Gemini AO image was transformed to the HST F814W coordinate frame yielding a pixel position of [676.08,307.13] on the WF3 chip.  Figure 1 shows the aligned pre- and post-explosion images, all of which are centred on the SN position. A source is visible at the SN site in the F450W, F606W and F814W pre-explosion images, although nothing is seen in the F300W frame (not shown in Figure 1).  Its position in the F814W image was measured using 5 different methods, DAOPHOT PSF fitting, {\sc hstphot} and the three centring algorithms within DAOPHOT - centroid, gaussian and ofilter. The mean pixel position was [676.11,307.17] with an uncertainty of $\pm$12 mas. With a measured displacement of 5 mas and a total astrometric uncertainty of $\pm$25 mas, this source can be considered coincident with SN 2008ax. This is the same object as suggested by \citet{li08a} to be the possible progenitor. However, high resolution images are necessary to reduce the positional error, and ambiguity, as much as possible.  

A slight pixel offset between the F450W and F814W images as measured by the {\sc Iraf} task {\it crosscor} is consistent with the offset in the position of the proposed progenitor, confirming that this is the same source in both filters. The F606W image was taken $\sim$7 years earlier with a pointing entirely different from the F450W/F814W observations. Positions of 35 point sources common to the F606W and F814W images were input to {\it geomap}, resulting in a transformation with an rms error of $\pm$16 mas. The F814W object position was transformed to the F606W coordinate frame and found to be coincident within the astrometric uncertainties. 

{\sc hstphot} flight system magnitudes of the pre-explosion source were measured as F450W = $23.66\pm0.10$, F606W = $23.36\pm0.10$, F814W = $22.63\pm0.10$. From pixel statistics at the SN position we estimate a 3$\sigma$ detection limit of F300W=22.9.
 

\begin{figure*}
\label{pre-post}
\includegraphics[width=17.5cm]{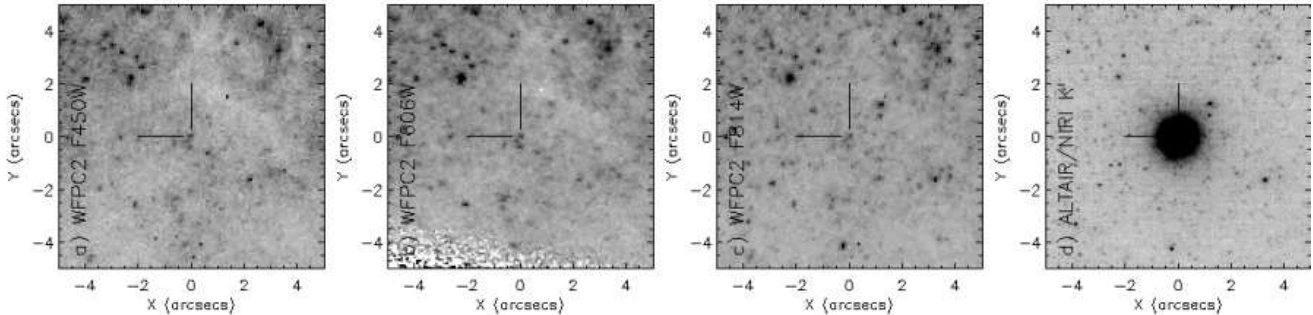}
\caption{Pre- and post-explosion images of the site of SN 2008ax in NGC 4490. Each frame is centred on the SN position and oriented such that North is up and East is to the left.  The position of SN 2008ax is indicated by the cross hairs. (a) Pre-explosion HST+WFPC2 F450W image. (b) Pre-explosion HST+WFPC2 F606W
image. (c) Pre-explosion HST+WFPC2 F814W image. (d) Post-explosion Gemini+Altair/NIRI $K$-band image in which the SN is clearly visible. An object coincident with the SN position is visible in all three pre-explosion frames. The HST+WFPC2 F300W pre-explosion image, in which this object is not detected, is not shown here.}
\end{figure*}

\section{Distance, reddening and metallicity}
\label{dist_red_metal}

Estimates of the distance to NGC 4490 and the reddening towards the SN progenitor are taken from a companion paper on the evolution of SN 2008ax \citet{past08}. Pastorello et al. derive a mean distance modulus of $\mu = 29.92 \pm 0.29$ and a reddening of $E(B-V)=0.3$ estimated from the equivalent width (EW) of the Na I doublet and the relations of \citet{tura03}. We note that \citet{chorn08} find a colour-excess of $E(B-V)=0.5$ despite measuring the same EW as Pastorello et al. 

A recent study has estimated the oxygen abundances of NGC 4490 from SDSS DR5 spectra of 5 individual \Hii regions \citep{pil07}, none of which are close to the SN position. 
Values between 8.3 - 8.5\,dex (on the scale of $12 + \log {\rm O/H}$) are measured around $R_{\rm G}/R_{\rm 25}  \sim 0.2$, while SN 2008ax is at a galactocentric distance of 0.3. Pilyugin \& Thuan calibrations give abundances which are slightly, but systematically, lower than previous studies. The H{\sc ii} region abundances measured by these authors at the position of SN 2008ax appear similar to those they measure in the central regions of M101, for which the best recent estimates suggest solar-like values \citep[e.g.,][]{bres07}. Hence we conclude that the progenitor of SN 2008ax most likely had solar-type metallicity and is bracketed from below by LMC-like values (i.e. 8.3\,dex) taking the Pilyugin \& Thuan data at face value.

\section{The Progenitor of SN 2008ax}

These pre-explosion images constitute the most detailed information on a SN progenitor since the identification of the precursor of SN~2003gd \citep{smartt04}, and the first for a peculiar type IIb SN since SN~1993J.  In the case of SN~1993J the progenitor was identified as a K-type supergiant with an UV and B-band excess interpreted as a nearby hot, massive early B-type star \citep{ald94,maund04}. Interaction with this proposed binary companion is used to explain the stripped nature of the SN progenitor. \citet{ryd06} found a point source at the site of the type IIb SN 2001ig $\sim$1000 d post-explosion which they identify as the blue supergiant companion to the putative progenitor. Given the type IIb classification of SN 2008ax one might assume a similar binary progenitor system to be appropriate. Below we compare the pre-explosion photometry and magnitude limits with several progenitor scenarios. Using extinctions derived with the $A_x/E(B-V)$ relations of \citet{vandyk99}, we find absolute photometry for the progenitor of $M_{F450W}=-7.55\pm0.31$, $M_{F606W}=-7.43\pm0.31$, $M_{F814W}=-7.86\pm0.31$ (Poisson noise = $\pm$0.10, $\mu_{err}$ = $\pm$0.29) and a $3\sigma$ detection limit of $M_{F300W}$=-8.6.\\


\noindent{\bf \emph{Single massive supergiant progenitor}}. The spectral energy distibutions of supergiant stars from \citet{drill00} were converted from UBVRI to HST flight system magnitudes using the colour corrections of \citet{maundsmartt05}. Comparing these to the observerd SED of the pre-explosion source we found no single star solution, in agreement with \citet{li08b}. This is not surprising given that the source is simultaneously blue in (F450W-F606W) and red in (F606W-F814W). Even when applying arbitrary levels of extinction we were unable to fit a single supergiant SED to the observed photometry.


\noindent{\bf \emph{Superposition of two supergiant stars}}. That SN 2008ax transitioned from a type II to a type Ib SN \citep{past08} implies that the progenitor had lost all but a small fraction ($\leq$ few $10^{-1}\rm M_{\odot}$) of its H-rich envelope. Previous studies suggest that the type IIb SNe 1993J and 2001ig arose from massive binary progenitor systems \citep{ald94,maund04,ryd06}, the binary orbit in each case being close enough to allow interaction. For SN 2008ax we do not restrict ourselves to an interacting binary scenario. The PSF of the proposed progenitor is consistent with that of a single star; however, its FWHM is some 0.15$\arcsec$, which translates to $\sim6$pc at the distance of NGC 4490. The source might actually be two stars separated by as much as 3 pc and such widely separated objects would not interact, evolving instead as two single stars. The progenitor in this case would have to lose almost its entire hydrogen envelope through strong stellar winds prior to exploding, a feat which theory suggests is only possible for stars more massive than $\sim$25-30 $M_{\odot}$ \citep{eldtout04}.

Pairs of supergiant SEDs from \citet{drill00}, converted to the HST VEGAMAG system (see above), were fitted to the source photometry with the same value of extinction being assumed for both components. The best fit was a B8 supergiant of $log(L/L_{\odot})\approx$5.15 combined with a M4 supergiant of $log(L/L_{\odot})\approx$4.20.  This implies the initial mass of the more evolved M-type supergiant ($\sim8-10 M_{\odot}$) is much less than that of its B-type companion ($\sim25 M_{\odot}$). Assuming the stars are coeval one would expect the more massive object to be the most evolved. Even if one argues that the more massive object is evolved, the lower mass star should not be observed as a red supergiant as its main sequence lifetime is $\sim$4-5 times longer. We also attempted to fit a Wolf-Rayet (WR) + supergiant interacting binary \citep[e.g.,][]{ryd04} using model WR colours (see discussion of single Wolf-Rayet progenitor for details) and found a similar result; the WR star was around 50 times more luminous than its evolved red/yellow supergiant companion.  Neither of these two-star models is self-consistent. 

The extinction towards an assumed companion star was varied, while retaining \mbox{$E(B-V)=0.3$} for the progenitor. Since this implies that the stars are separated by some considerable distance, we cannot invoke binary interaction. Arbitrarily reducing the companion extinction to zero we found a range of fits where the components had quite similar luminosities/masses, the progenitor being of later spectral type. The relative evolutionary timescales are not inconsistent in this case; however the implied main sequence mass of 10-14 $M_{\odot}$ is much too low for the progenitor to have lost most of its hydrogen envelope through wind-driven mass-loss. 

It is possible that several other stars might lie within the PSF as was the case for the progenitor of SN 1993J \citep{ald94,maund04}. Removing the flux contributed by these stars might allow us to fit a consistent interacting binary model. 


\noindent{\bf \emph{Young stellar cluster}}. Alternatively the source might be an unresolved, young stellar cluster in which the SN progenitor was embedded, provided the cluster diameter is less than $\sim$6 pc (PSF FWHM). This is not unreasonable as compact stellar clusters observed in M51 have median effective radii of 2-4 pc \citep{lee05,scheep07}. Previous studies have attempted to place constraints on SN progenitors by estimating the ages and main sequence turn-off masses of their host clusters (e.g., SN 2004dj - \citealt{maiz04a}; SN 2007gr - \citealt{rmc08}; GRB 030329/SN 2003dh - \citealt{ost08}). Here we attempt a similar analysis assuming that the pre-explosion source is a compact, coeval cluster. \citet{bast05} suggest sources brighter than $M_V$=-8.6 are most likely star clusters. Our source is somewhat less luminous ($M_{F606W}=-7.43$) so cannot immediately be characterised as such. We used CHORIZOS 
(version 2.1.4) \citep{maiz04b}, a $\chi^2$ minimisation code, to fit Starburst99 \citep{leit99} stellar populations to our observed photometry.  The model SEDs were of solar metallicity and assumed a Salpeter initial mass function with an upper mass cutoff of $100\rm M_{\odot}$. CHORIZOS solved for cluster age and extinction using a \citet{card89} extinction law with $R_V$=3.1. The best fit was a cluster with an age of $\sim$20 Myr (turn-off mass $\sim12\rm M_{\odot}$), a total mass of $\sim1800\rm M_{\odot}$ and extinction of \mbox{$E(B-V)=0.15$}, although a range of almost equally significant solutions was found with ages of 8-28 Myr and \mbox{$E(B-V)=0.0 - 0.2$}. This degeneracy was due to the lack of information at UV and NIR wavelengths. 
Since none of these `best fits' had an extinction consistent with our measured value, we forced CHORIZOS to adopt \mbox{$E(B-V)=0.3$}. The solution in this case was a 42 Myr cluster (turn-off mass $\sim8\rm M_{\odot}$)
; however, the model SED was under-luminous in $F450W$ and over-luminous in $F606W$ compared with the observed photometry. 

Given the relatively low mass estimates for these clusters, it is perhaps not surprising that we do not find a satisfactory match between the models and our observed SED. The model SEDs assume an infinite number of stars in the cluster and therefore a fully populated IMF; a reasonable approximation for more massive clusters. For low mass clusters, where the IMF is more sparsely populated and the total luminosity is much lower, even a single massive star can significantly affect the observed SED; and the stochastic nature of the masses of the stars, especially the most massive stars, can result in a large dispersion in the integrated photometry of clusters of similar masses \citep[e.g.,][]{cervino04,jamet04}. In an attempt to mimic these effects we created a grid of `cluster+massive star' models by combining the flux from a single, evolved star (of mass 10-30 M$_{\odot}$ and spectral type WR/O9-M5) with coeval ($\sim$6 to 27 Myr) Starburst99 stellar populations. The single massive star was assumed to be the SN progenitor. These hybrid models were fitted to the observed photometry adopting a value of \mbox{$E(B-V)=0.3$}. The resulting fits yielded cluster masses of $\sim1500-3000\rm M_{\odot}$, main-sequence masses for the progenitor of 10-14 M$_{\odot}$, and colours consistent with OB-type stars and WR objects that have retained a fraction of their H envelope. Stars of such low initial mass could not have produced a H-poor progenitor through wind-driven mass-loss. Hence we conclude that, if embedded in a cluster, the progenitor star must have been stripped in an interacting binary system. Interestingly these initial masses are similar to that estimated for the progenitor of SN~1993J ($\sim$15 M$_{\odot}$), although it exploded as a cooler, K-type star.

\begin{figure*}
\label{fig2}
\includegraphics[width=16cm]{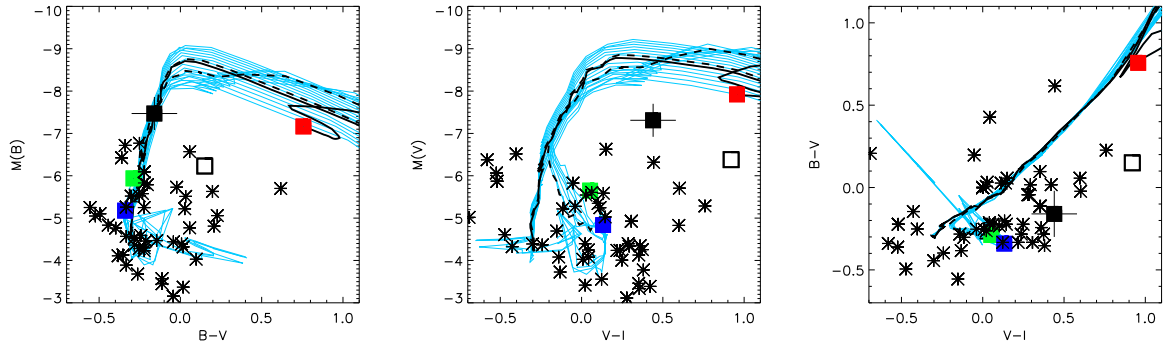}
\caption{Comparison of the pre-explosion source photometry with {\sc STARS} stellar evolution models \citep{eldvink06}. The blue lines are non-rotating solar metallicity models between 20-32M$_{\odot}$. The thick black line is the 27M$_{\odot}$ model; the dashed line the 28M$_{\odot}$ model. The red, green and blue squares indicate the endpoints of the 27, 28 and 29M$_{\odot}$ models respectively. The black asterisks are observed WN and WNL stars in M31 and the LMC \citep{massey05,massey02}, which have been corrected for Galactic extinction. The possible SN progenitor is marked by the black square, with the hollow square its position before correcting for extinction.}
\end{figure*}

\noindent{\bf \emph{Single Wolf-Rayet star}}. A progenitor star of sufficient initial mass could have lost most of its H-rich envelope through radiatively driven mass loss. WR stars, exposed He cores of massive stars, can form in this way. Helium-rich WN stars have been proposed as possible progenitors of type Ib SNe and H-rich supergiants have been shown to be the progenitors of type II SNe \citep[e.g.,][]{smartt04,li06}. A progenitor intermediate of these types of star would be consistent with the strange evolution of a type IIb SN. 

We compared our pre-explosion photometry with solar metallicity models of single stars of initial masses between 20 and 32 $\rm M_{\odot}$. Our model colours and absolute magnitudes were obtained using the method of \citet{lej01} to perform synthetic photometry on the Cambridge {\sc STARS} stellar evolution models \citep{eldvink06}. This process is described in more detail by \citet{eld07}, although here we have also calculated synthetic photometry for WR stars using the model WR spectra of \citet{graf02} and \citet{ham03}. Photometry was calculated using both the Johnson-Cousins and HST WFPC2 photometric systems, so that it was possible to compare the models directly with BVI photometry of observed WR stars and WFPC2 photometry of the proposed progenitor respectively. Figure 2 shows BVI colour-magnitude and colour-colour plots of these models along with WN and WNL stars in M31 and the LMC \citep{massey05,massey02}. The photometry of the proposed progenitor, transformed to BVI magnitudes \citep{holtz95,dolphcalib}, is plotted with a solid square and its colours fit well with those of the observed WR stars. We note that the observed WR stars show a considerable dispersion in colour, particularly in V-I, when compared to the model tracks. The strong mass-loss and emission-line dominated spectra of these objects lead to an inherent spread in their broad-band colours \citep[e.g.,][]{ham95}. Given the ambiguous nature of the object's SED our transformation WFPC2 to Johnson-Cousins magnitudes maybe somewhat uncertain, but comparison of the flight-system magnitudes with the WFPC2 model photometry shows a very similar relationship between the pre-explosion source and the models to that shown in Figure 2. The progenitor lies beween the endpoints of the 27 and 28 M$_{\odot}$ tracks. Remarkably these model stars end their lives while transitioning between the supergiant and WR phases, precisely the type of progenitor one would expect to produce a type IIb SN. Final hydrogen masses are between $\sim10^{-1}$ and $10^{-3}$ M$_{\odot}$. These would be classed as WNL or WNH stars \citep{smith08}. 

A potential pitfall for our single WNL/WNH progenitor is the mass of its carbon/oxygen (C/O) core. The final stellar mass is  11-12 M$_{\odot}$, with a C/O core mass of $\sim$9 M$_{\odot}$; a factor of two higher than for the progenitor of SN 1993J \citep{maund04}. \citet{past08} show that the bolometric light curves and ejecta velocities of SNe 2008ax and 1993J are very similar, inferring that ejecta masses, and hence progenitor masses were also similar.  They suggest the more massive C/O core arising from our single star model would have produced a broader light curve. If these two progenitors actually had core masses and ejecta masses different by a factor of two, then the fact that they show similar evolution would have important implications for modelling of SN lightcurves. Alternatively, the ejecta masses could be similar if we assume our WNL/WNH progenitor produced a more massive compact remnant ($\sim$5 M$_{\odot}$) than was the case for SN 1993J. The remnant formed would be a black hole, in agreement with the predictions of stellar models in this initial mass range \citep{heger03,eldtout04}. 

\section{Conclusions}

We have determined the position of SN 2008ax to within $\pm$22 mas in pre-explosion HST observations of NGC 4490 and identified a coincident source in three filters. It is unclear whether this object is the SN progenitor alone, or a blend of it and several other stars. We conclude that it is not a single supergiant star, but that an OB/WR star progenitor (of main-sequence mass 10-14 M$_{\odot}$) in an interacting binary is possible so long as we include a significant flux contribution from an unresolved stellar cluster. We also find a single star progenitor model consistent with the pre-explosion photometry, where an initially very massive star ($\sim$28 M$_{\odot}$) loses most of its H-rich envelope prior to exploding as an 11-12 M$_{\odot}$ WNH star. However, the light curve of SN 2008ax would suggest a lower progenitor mass \citep{past08}. In several years time, when the SN has faded, further high resolution imaging may help us better determine the nature of the progenitor star. If a source is still visible, subtraction from the pre-explosion data could help to isolate the progenitor SED. If the source has disappeared, it could have serious implications for the interpretation of ejecta masses from SN lightcurve and spectral modelling.

\section*{Acknowledgments}
We thank Jes{\'{u}}s Ma{\'{i}}z-Apell{\'{a}}niz for helpful comments and discussion. Observations: NASA/ESA Hubble Space Telescope obtained from the Data Archive at the Space Telescope Science Institute; Gemini Observatory. We acknowledge funding through EURYI and ESF (SJS, RMC), NSF/NASA grants AST-0406740/NNG04GL00 (JRM), Academy of Finland project: 8120503 (SM), Benoziyo Center for Astrophysics and Eda Bess Novick New Scientists Fund at the Weizmann Institute of Science (AGY).

\end{document}